\documentclass{sig-alternate-hotpets}

\usepackage{graphicx}
\usepackage{url}
\usepackage{xspace}
\usepackage{enumitem}
\usepackage[font=small,labelfont=bf]{caption}

\usepackage{tikz}
\usetikzlibrary{positioning}
\usetikzlibrary{shapes,arrows}
\usetikzlibrary{calc}
\usetikzlibrary{fit}

\usepackage[cm]{sfmath}

\DeclareGraphicsExtensions{.png}
\usepackage{macros}

\begin{document}
\title{Extending the Anonymity of {\ttlit Zcash} }
%
%
%
%
%

\numberofauthors{2} 
%
\author{
%
%
\alignauthor
George Kappos\\
       \affaddr{University College London}\\
\alignauthor
Ania M. Piotrowska\\
       \affaddr{University College London}\\
}

\maketitle

\setlength{\parindent}{2em}

\section{Introduction}

Although Bitcoin~\cite{1} in its original whitepaper stated that it offers anonymous transactions,
de-anonymization techniques have found otherwise~\cite{2,4}.
Therefore, alternative cryptocurrencies, like Dash\footnote{\scriptsize\url{https://www.dash.org}}, Monero\footnote{\scriptsize\url{https://getmonero.org}} and \zcash\footnote{\scriptsize\url{https://z.cash}},
were developed to provide better privacy.
As Edward Snowden stated \emph{"Zcash's privacy tech makes it the most interesting Bitcoin alternative~[...] because the privacy properties of it are truly unique".}
\zcash's privacy is based on peer-reviewd cryptographic constructions,
hence it is considered to provide the foundations for the best anonymity.
However, even \zcash makes some privacy concessions.
It does not protect users' privacy in the presence of a global adversary who is able to
observe the whole network, and hence correlate the parties exchanging money, by using their network addresses.
The recent empirical analysis of
\zcash~\cite{17} shows, that users often choose naive ways while performing the protocol operations,
not realising that it degrades their anonymity.

In this talk, we will discuss an extension of
\zcash using mix networks to enhance the privacy guarantees of users that choose to remain anonymous by tackling
two major security challenges: one at the application layer of the scheme and one at its network layer.


\section{zcash in a nutshell}

\zcash offers two types of user addresses, the \taddr, which is also used in Bitcoin, and the \zaddr,
which has the purpose of hiding the identity of its owner.
\zcash inherits Bitcoin's functionality,
in the sense that the sender may choose to perform a transparent transaction (\TT),
in which both the spender and the recipient of the coins are identified by a \taddr. However, transparent transactions enable
tracking the whole transaction history of any given coin.
In order to break the link between senders and recipients,
\zcash enables a user to transact privately, either through a \emph{shielded} transaction (\TZ) in which the recipient's address remains hidden, a \emph{deshielded} (\ZT) one in which the
sender's address remains hidden; or a \emph{private} (\ZZ) one, in which both
of the addresses are unknown and the value of the spent coin remains secret.

\begin{figure}[t!]
    \centering
    \resizebox{\columnwidth}{!}{
    \begin{tikzpicture}[font=\sffamily]
        \node (pool) [draw, fill=gray, opacity=.3, rounded corners, thick, shape=rectangle, minimum width=2cm, minimum height=3.6cm, anchor=center, label={\scriptsize POOL}] at (0,0) {};

        \node (z1) [on grid, draw, circle, thick, below left=0.23cm of pool.north] {$\scriptstyle z_1$};

        \node (z4) [on grid, draw, circle, thick, below=0.96cm of z1.west] {$\scriptstyle z_4$};

        \node (z2) [on grid, draw, circle, thick, below right=0.8cm of z1] {$\scriptstyle z_2$};

		\node (z5) [on grid, draw, circle, thick, below=1.48cm of z2] {$\scriptstyle z5$};

		\node (z6) [on grid, draw, circle, thick, below=0.5cm of z5.east] {$\scriptstyle z6$};

		\node (z3) [on grid, draw, circle, thick, below=0.4cm of z2.east] {$\scriptstyle z_3$};

        \node (t4) [draw, rounded corners, thick, shape=rectangle, minimum width=0.8cm, minimum height=0.8cm, minimum size=0.8cm, right=1.6cm of z5] {$\small t_4,X_1$};

		\node (t1) [draw, rounded corners, thick, shape=rectangle, minimum width=0.8cm, minimum height=0.8cm, minimum size=0.8cm, left=1.2cm of z1] {$\small t_1,X_1$};

		\node (M) [draw, rounded corners, thick, shape=rectangle, minimum width=0.95cm, minimum height=0.85cm, left=1.8cm of z5] {$ Mix $};

		\node (t2) [draw, rounded corners, thick, shape=rectangle, minimum width=0.8cm, minimum height=0.8cm, minimum size=0.8cm, above=1.2cm of t4] {$\small t_2,X_1$};

		\node (t5) [draw, rounded corners, thick, shape=rectangle, minimum width=0.8cm, minimum height=0.8cm, minimum size=0.8cm, below=0 of t4] {$\small t_5,X_2$};

		\node (t3) [draw, rounded corners, thick, shape=rectangle, minimum width=0.8cm, minimum height=0.8cm, minimum size=0.8cm, left=0.8cm of M] {$\small t_3,X_3$};

        \node (p2p) [draw,dotted,thick, rounded corners, inner sep=8pt, fit=(t1) (pool) (t2), label={[shift={(-2.4,0.0)}]\small P2P NETWORK}] {};

		 \path [->, draw, thick] (t3) -- node [above] {\footnotesize com} (M);

		 \path [->, draw, thick] (M) -- (t3);

		 \node (lan) [draw, dotted, thick, rounded corners, inner sep=10pt, fit=(M) (t3) ] {};

		 \node (adv) [draw, dashed, color=red, thick, rounded corners, inner
		 sep=12pt, fit=(p2p) (lan), label={[shift={(-4,0.0)}]\small ADVERSARY}] {};

		 \path [->, draw, ultra thick] (z1) --  node [below,sloped] {$\scriptstyle TX_2, X_1 ZEC$}  coordinate(z1-t2) (t2);

		 \path [->, draw, ultra thick] (t1) --  node [above] {$\scriptstyle TX_1, X_1 ZEC$}  coordinate(t1-z1) (z1);

		 \path [->, draw, ultra thick] (t1) --  node [above,sloped] {$\scriptstyle TX_1, 0 ZEC$}  coordinate(t1-z4) (z4);

        \path [->, draw, ultra thick] (M) --  node [above] {$\scriptstyle TX_3, X_1 ZEC$} coordinate(M-z5) (z5);

        \path [->, draw, ultra thick] (M) --  node [below,sloped] {$\scriptstyle TX_3, X_2 ZEC$} coordinate(M-z6) (z6);

        \path [->, draw, ultra thick] (z5) -- node [above] {$\scriptstyle TX_4, X_1 ZEC$} coordinate(z5-t4) (t4);

        \path [->, draw, ultra thick] (z6) -- node [above] {$\scriptstyle TX_5, X_2 ZEC$} coordinate(z6-t5) (t5);

    \end{tikzpicture}
    }
\caption{\scriptsize Naive usage of the Zcash pool VS our implementation using a Mix Network. The set of z-addresses visualised as a centralised pool, for ease of understanding.}
\label{fig:Fig1}
\end{figure}
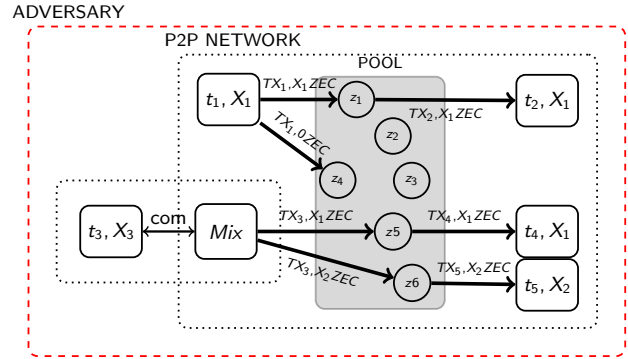

\subsubsection*{\large{\textbf{Attacks on Zcash's anonymity}}}
Although applied to different layers, both of the attacks we describe below have a common goal:
To correlate a \TZ transaction with a \ZT one, hence to identify the two
end $t$-addresses that are involved in this sequence of transactions. The first attack, at the application
layer, can be performed by anyone with access to  the \zcash blockchain, whereas the second
one, at the network layer, can be applied by a global passive adversary (GPA) who observes the whole \zcash P2P network.

\par \noindent \textbf{Application Layer:}
In a \TZ transaction, \zcash specifies that the
sender must select two distinct destination $z$ addresses, like $z_1$ and $z_4$ in Figure~\ref{fig:Fig1},
to give the incentive to the user to split the coin into two coins and avoid
a withdrawal of the same value that was earlier deposited. However, as the analysis
in~\cite{17} showed, most users send the entire coin value to $z_1$ and $0$ $ZEC$
to $z_4$, as shown in $TX_1$ in Figure~\ref{fig:Fig1}. This allows the adversary to correlate the transactions $TX_1$ and $TX_2$.

\noindent \textbf{Network Layer:}
As recent revelations show\footnote{\scriptsize\url{https://theintercept.com/2018/03/20/the-nsa-worked-to-track-down-bitcoin-users-snowden-documents-reveal/}}, Bitcoin is under an extensive surveillance from the NSA, which monitors its blockchain and global traffic in order to identify users and their transactions. This shows that a GPA capable of observing the P2P network is not a theoretical bugbear, but a realistic danger. Therefore, it is crucial to secure the cryptocurrency at its network level and protect the privacy of the users.
A GPA who is able to observe the  Zcash network can easily discover the network
addresses of users who broadcast, hence correlate their transactions~\cite{12}.
This attack is applicable to most of the deployed cryptocurrencies. Tor\footnote{\scriptsize\url{https://www.torproject.org}} has been suggested as a possible defence, however, as shown in~\cite{12} it is not an ultimate solution since even a low-resourced adversary can perform attacks. These not only deanonymize the participants of the transaction but more importantly allow the adversary to
control which blockchain state is visible to the users or which transactions
are relayed to them. Moreover, Tor is not resistant to a GPA, therefore it is not suitable for our threat model.

\section{Mixes for anonymity}

A mix network~\cite{15} is a sequence of cryptographic relays, that hides network level metadata,
by using end-to-end layered encryption and secret mixing of packets.
The more packets are mixed together, the better
anonymity. Therefore, traditional mix networks used long batching times and cover traffic, thus significantly limiting their usage
to only high-latency communication.
Recent research shows~\cite{16} that it is possible to build low-latency
mix networks by using tunable cover traffic and delays.
In contrast to Tor, mix networks protect the users' privacy even in the presence of a GPA performing sophisticated traffic analysis.
In this talk, we want to discuss the idea of using mix networks as the proxy between the users and the \zcash network.
Implementing a mix network over \zcash would ensure that users' transactions remain anonymous
even if the adversary is able to observe the communication channel between the users' local networks and the whole P2P network.
Merging mix networks with cryptocurrencies has a bilateral benefit. On one hand, it allows for truly
anonymous transactions. On the other hand, it offers another use-case for mix networks, thus increases their potential.

\section{Our scheme}

In our proposal, we leverage a mix network as a proxy channel between the users and the \zcash network.
The mix network would serve the two following roles:

\phead{Broadcast}
It would be responsible for broadcasting all of the users' transactions to the P2P network.
The user encapsulates the transaction into the cryptographic packet format and passes it to the first mix server.
Thanks to that, only the first server in the mix chain knows the
network address of the user and every other server learns only the address of the previous and the next server.
The last server in the chain is responsible for broadcasting the decrypted transaction into the Zcash network.
Hence, the network layer attack is now impossible since the network address of the user is only visible to the GPA in the user's network
and it cannot be correlated with any transaction in the blockchain, since the encryption of the packet hides its source,
and the broadcasting address belonging to the mix server.

\phead{Suggest coin splitting}
We suggest using mix servers as advisors who recommend to the users
the optimal split of their coins, hence help them protect themselves against application
layer attacks. Mix servers can aggregate information about prior deposits into the pool and
based on this knowledge calculate the best strategy for splitting the coins, which increases the anonymity set of the users. Such blockchain analysis is very low-cost and can be performed by modern personal computers.
For example, Figure~\ref{fig:Fig1} shows how $Mix$ takes into consideration the transaction $TX_1$ that happened earlier
by recommending to the user $t_3$ to break his coin $X_3$ as $X_3=X_1+X_2$.


\section{Discussion Points}

The primary goal of this talk is to hear the community's feedback and suggestions about
the scheme, since such design raises many challenges, as well as to spur the discussion regarding the anonymity of the cryptocurrencies.
We intend to debate the following points:
\begin{itemize}[noitemsep, leftmargin=*, topsep=0pt]
    \item
        Should the Zcash client enforce the users to transact anonymously by making the private transactions the only option, or should the choice remain? An interesting discussion, as well, is whether and how we can give the users the incentive to use the pool.
    \item In order to offer the anonymity properties, a mix network uses additional delays, hence
        it would introduce additional latency overhead for the transactions. In our opinion,
        a thoroughly evaluated per hop delay does not significantly increase the overall latency,
        hence does not degrade the user experience.
    \item Some of the mix nodes can be malicious. One of the possible attacks that a corrupt mix can perform
        is a DoS attack, in which the mix never transmits the received packets, hence decreases the system's reliability.
        One possible solution would be for the user to send the same packet carrying a transaction to multiple mix cascades.
        More importantly, the malicious mix servers cannot craft their own transactions in order to steal users' money.
    \item In order to ensure that the shielded pool processes a large enough number of transactions, hence guarantees
        anonymity, and that the density of traffic in the mix net does not leak any communication information,
        we have to use cover traffic. The question of how much of such cover traffic is required opens up a new research challenge which we aim to study.
    \item What are the requirements for such design in the opinion of the research and developers communities.
\end{itemize}



\bibliographystyle{abbrv}
\bibliography{bibli}  
%

\balancecolumns
\end{document}